\documentstyle[11pt,epsfig,psfrag,twoside]{article}
\newcommand{\bea}{\begin{eqnarray}}
\newcommand{\eea}{\end{eqnarray}}
\newcommand{\simgt}{\hbox{ \raise3pt\hbox to 0pt{$>$}\raise-3pt\hbox{$\sim$} }}
\newcommand{\simlt}{\hbox{ \raise3pt\hbox to 0pt{$<$}\raise-3pt\hbox{$\sim$} }}

\begin{document}
\setcounter{page}{1}
\title{Top quark and QCD physics at $e^+e^-$ linear colliders:
recent progress\thanks{Talk given 
at the International Linear Collider Workshop (LCWS02), 
Jeju island, Korea, Aug.\ 2002.}} 
\author{Y.~Sumino
\\
\\
{\it Department of Physics, Tohoku University,
Sendai, 980-8578 Japan}
}
\date{}
\maketitle
\begin{abstract}
\noindent
I review the studies, which were reported after
the last Linear Collider Workshop, on top quark physics and QCD physics
at a future $e^+e^-$ linear collider.
\end{abstract}

\section{Introduction}

Since the first Linear Collider (LC) Workshop, an enormous amount
of studies have been done on top quark physics and
QCD physics that can be covered at a future linear collider.
At early stages, many new ideas were proposed.
More recently, studies are centered toward precise theoretical
predictions and simulation studies incorporating more realistic
and advanced experimental setups.
It indicates a healthy direction of the
development, since the role of the top quark
and QCD physics within the grand aim of linear collider experiments
is to search for new physics through precision
studies of top quark properties and through accurate understanding
of strong interaction phenomena.
It is quite impressive to see
how interesting and important progress is still being made.

In this article I summarize the works presented in the 
top/QCD parallel session at LCWS02.
In addition I also summarize some of the important
works which appeared
after the last LC Workshop.
The following subjects are reviewed:
an updated analysis on parameter determinations at $t\bar{t}$
threshold;
an improvement of theoretical predictions for the top threshold
cross section;
the role of top quark offshellness in the toponium energy level;
a new decoupling theorem for the top decay form factors;
a kinematical fitting method for top event reconstructions;
jet momentum distributions in a flux-tube model.
I apologize in advance to those, whose works are not given full
justice in my summary.

\section{Parameter determinations at top threshold: update}

An updated parameter determination study has been carried out
in the top quark threshold region \cite{Martinez-Miquel},
which includes the following new aspects:
(1) It is based on simultaneous measurements of three physical
observables, the top production cross section
$\sigma_{t\bar{t}}$, the top momentum $p_t$,
and the forward-backward asymmetry of the top quark $A^t_{\rm FB}$,
which are extracted from the same top quark sample.
(2) It performs multiparameter fits (up to four parameters:
the top quark $1S$-mass $m_t(1S)$,
the strong coupling constant $\alpha_S(M_Z)$,
the top width $\Gamma_t$,
and the top-Higgs Yukawa coupling $g_{tH}$.
(3) Systematic uncertainties are partly included.

\begin{figure}
\epsfxsize=7cm
\leavevmode
\centering
\epsffile{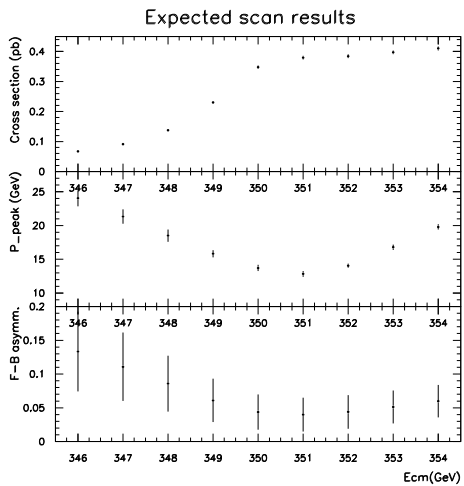}
\caption{\small
The expected scan results for the cross section, the peak of 
the top momentum distribution and the forward-backward charge asymmetry
\cite{Martinez-Miquel}.
}
\label{fig:Experiment}
\end{figure} 

Fig.~\ref{fig:Experiment} 
shows the energy dependences of the three observables in
the threshold region together with estimated errors corresponding to 
an integrated luminosity of 300~fb$^{-1}$.
The analysis showed that 
(a) $\sigma_{t\bar{t}}$ is most important for
the determination of the above mentioned parameters,
(b) $p_t$ reduces correlations of the errors in the parameter
determinations considerably,
(c) $A^t_{\rm FB}$ plays a rather minor role.

For instance, with a 300~fb$^{-1}$ integrated luminosity, 
a 3-parameter fit was performed
and resulted in the following uncertainties of the
top mass, strong coupling and top width:
\bea
\Delta m_t(1S) = 19~{\rm GeV},
~~~
\Delta \alpha_S(M_Z)=0.0012,
~~~
\Delta \Gamma_t = 32~{\rm MeV} .
\eea
Particularly the accuracies estimated for $m_t(1S)$ and $\Gamma_t$
are quite impressive:
$10^{-4}$ and 2\% relative accuracies, respectively.
Nevertheless, one should note that presently there remain
theoretical uncertainties of order 100~MeV in relating
the $1S$-mass $m_t(1S)$ to the $\overline{\rm MS}$-mass, the latter of
which we would like to determine eventually \cite{topcollab}.
Thus, the above accuracy of $m_t(1S)$ should be regarded as
an accuracy goal with which theorists should predict
the relation between the $1S$-mass and 
the $\overline{\rm MS}$-mass in the future.

The above errors agree (roughly) with those which can
be obtained by scaling the error estimates of previous analyses 
(e.g.\ \cite{fms})
by square-root of the integrated luminosity.
However, the non-trivial point is that in this new analysis
a 3-parameter fit was performed, whereas in the previous
analyses only one or two parameters were varied while
fixing the others.
That there is no significant loss of accuracies when the multiparameter
fit is performed, shows the smallness of the error correlation 
thanks to the simultaneous measurements of the 
physical observables.

A 4-parameter fit was also performed including the top Yukawa
coupling in addition, for $M_H=120$~GeV.
The analysis showed that the determination of the
top Yukawa coupling is quite challenging
in the top threshold region.

\section{Improved predicton for $\sigma_{t\bar{t}}$
in the threshold region}

In the last two LC Workshops, it was recognized that
there exist large uncertainties in the present theoretical
prediction for the normalization of the top quark production
cross section $\sigma_{t\bar{t}}$ in the threshold region.
In order to improve accuracy of the theoretical prediction, 
Ref.~\cite{rg} included resummation of logarithms
$\alpha_S \times (\alpha_S \log \alpha_S )^n$
into the cross section.
This prescription stabilizes the prediction substantially, and 
it is claimed that the 3\% theoretical
accuracy is achieved for the normalization of $\sigma_{t\bar{t}}$.

Up to now, a full consensus has not been reached among theorists on this
3\% theoretical accuracy.
In particular, I find the following two questions relevant:
(1) Since $\log \alpha_S$ is not
particularly large, if the resummation of logarithms results in a
significant effect, one expects that non-logarithmic
terms may also contribute with a similar magnitude.
Indeed significance of the non-logarithmic term in 
the next-to-next-to-next-to-leading order (NNNLO) correction
has been suggested in \cite{Penin-Steinhauser,ks3}.
(2) Even if the resummation of logarithms 
is shown to stabilize the theoretical prediction
for the threshold cross section, what is the physical meaning
behind it?
Answers to these questions may be given by a rather different type of
consideration, which is discussed below.

\section{Offshellness of $t$,$\bar{t}$ as an IR cutoff}

Let me state briefly a historical background related to this analysis.
It is known that the sum of the pole masses of $t$ and $\bar{t}$
and the QCD potential between them,
$2 m_{\rm pole} + V_{\rm QCD}(r)$, contains an
${\cal O}(\Lambda_{\rm QCD}^3)$ perturbative uncertainty.
Within the potential-NRQCD framework, this uncertainty is absorbed
into a non-local gluon condensate \cite{BraPineSotVai}.
On the other hand, it has been known for a long time that the
leading non-perturbative corrections 
to the resonance energy levels are of order
$\Lambda_{\rm QCD}^4$ because it is proportional
to the local gluon condensate $\langle G_{\mu \nu} G^{\mu\nu} \rangle$
\cite{vol}.
Thus, there is an apparent mismatch in the power of $\Lambda_{\rm QCD}$
between the two quantities.

It was shown \cite{offshell} that when the offshellness of
$t$ and $\bar{t}$ is incorporated properly, it 
provides an additional suppression factor
of order $\Lambda_{\rm QCD}/(\alpha_S^2 m_t)$ to the 
perturbative uncertainty of the resonance
energy levels:
\bea
\delta E_n \sim \Lambda_{\rm QCD} \times 
\frac{\Lambda_{\rm QCD}^{\,2}}{(\alpha_S m_t)^2} \times
\framebox[14mm]{
${\displaystyle \frac{\Lambda_{\rm QCD}}{\alpha_S^2 m_t}}$
} .
\eea
Hence, the dimension of the perturbative uncertainty becomes
the same as that of the leading non-perturbative correction.
Qualitatively the suppression mechanism can be understood as follows.
If the offshellness is larger than $\Lambda_{\rm QCD}$, 
the rescattering time 
$\Delta t \sim (\alpha_S^2 m_t)^{-1}$ of $t$ and
$\bar{t}$ inside the resonance state becomes shorter than the
hadronization time scale.
That is, $t$ and $\bar{t}$ will get distorted before 
infrared gluons surround them, so that the offshellness acts
as an infrared cutoff of the temporal dimension.

One may recall that a large uncertainty 
of order $\Lambda_{\rm QCD}$ was
inherent in the theoretical prediction for the resonance energy levels
when the top quark pole mass
had been used conventionally.
The uncertainty was suppressed by 
$\Lambda_{\rm QCD}^{2}/(\alpha_S m_t)^2$ when a
top quark short-distance mass (such as the $\overline{\rm MS}$ mass)
was used instead.
This corresponds to an incorporation of the infrared cutoff of the
spacial dimension, namely the fact that the physical size of the
$t\bar{t}$ resonance state is much smaller than the typical hadron size.
Furthermore additional suppression factor comes from the
infrared cutoff in the temporal extent.
Correspondingly the series expansions of the energy levels become
more convergent when these effects are incorporated, and
we may achieve more accurate theoretical predictions.

Conceptually the suppression mechanism by offshellness is clear. 
It is, however, a non-trivial task to incorporate the offshell
suppression effects into theoretical
predictions systematically.\footnote{
In the usual approach to non-relativistic boundstates,
one starts from instantaneous gluon exchange in the leading order
and incorporates perturbative corrections to it.
Although this is justified for the leading binding kinematics,
such treatment 
spoils the role of offshellness as an infrared cutoff.
}
It has already been shown that these effects
include (as a part of the effects) resummation of 
ultrasoft logarithms.
Hence, it may provide answers to the questions
raised in the previous section, although more detailed analyses
are still needed.

Closely related to this subject,
we note the important progress in
the theoretical prediction for $e^+e^- \to t\bar{t}$ in the threshold region.
Theorists are now aiming at calculations of NNNLO corrections to the
cross sections.
Since the last LC Workshop, important computations
have been achieved toward this goal \cite{nnnlo-H,Penin-Steinhauser}.

\section{``Decoupling theorem'' for top decay form factors}

Related to the subject of top quark form factor determinations,
a new ``decoupling theorem'' was found \cite{new}.
The theorem can be phrased as follows.
Consider a process where two initial particles 
(``1'' and ``2'') collide and the top quark is produced
in association with some other particle(s) and the top quark
decays semi-leptonically:
$
1 + 2 \to t + X \to b \ell \nu + X .
$
In this process we assume presence of general form factors in the
top quark decay vertex.
There are four of them, $f_{1,2}^{L,R}$ and $\bar{f}_{1,2}^{L,R}$:
\bea
\Gamma^\mu_{Wtb} = - \frac{g}{\sqrt{2}} V_{tb} \,
\bar{u}(p_b) \biggl[
\gamma^\mu ( f_1^L P_L + f_1^R P_R ) 
-
\frac{i\,\sigma^\mu_{~\nu}k^\nu_W}{M_W}
( f_2^L P_L + f_2^R P_R )
\biggr] u(p_t) ,
\eea
where $P_{L,R} = (1 \mp \gamma_5)/2$.
At tree level of the Standard Model (SM), $f_1^L  =1$, and
all the other form factors are zero.
The other parts of the amplitude can be of any form
(e.g.\ anomalous form factors in the top production
vertex may be included),
as long as it includes the top quark as an
intermediate state which has the above decay vertex.
Then, the theorem states that the angular distribution of the
lepton $\ell$ from the top quark, defined in the laboratory frame, 
is independent of the anomalous form factors.
Here, the anomalous form factors stand for the deviations of the form
factors from the tree-level SM values.
This theorem holds under fairly general conditions:
\begin{itemize}
\vspace{-2mm}
\item
The initial particles (``1'' and ``2'') can have longitudinal momenta.
Hence, the theorem is applicable also to processes at a
photon-photon collider or at a hadron collider.
\vspace{-2mm}
\item
The mass of $\ell$ is neglected, but the bottom quark mass can be
non-zero.
\vspace{-2mm}
\item
Narrow width limits are assumed:
$\Gamma_t/m_t,~\Gamma_W/M_W \ll 1 .$
\vspace{-2mm}
\item
The theorem is valid up to linear terms in the anomalous
form factors, i.e.\ the terms quadratic in the anomalous form factors
are neglected.
\end{itemize}

\vspace{-2mm}
In measurements of the top quark form factors at $e^+e^-$ collider,
this theorem provides a useful tool for disentangling the
top quark form factors associated with the top quark production and
the decay vertices.
In fact if we first analyze the angular distribution of leptons 
from the top quark, this is sensitive only to the top quark production
vertex.
After determining the form factors in the production vertex, we can
use other observables for extracting the decay form factors.
Physically this theorem guarantees that the $\ell$ angular distribution
is an ideal analyzer of the top quark spin.
This fact has been known within the SM \cite{jk1};
the above theorem extends this picture to the case where there are
small anomalous form factors in the top quark interactions.
Since the top quark spin plays a crucial role in the analysis of
various top quark properties, this observation can be utilized
in many ways.
e.g. This theorem has been applied to the analysis of the
$CP$ property of the Higgs boson using the process
$\gamma \gamma \to H \to t\bar{t}$ \cite{gds}.

\section{Kinematical reconstruction in top threshold region}

Measurements of the top quark form factors will be carried out
both in the open top region and in the top threshold region.
So far most of the theoretical analyses as well as simulation 
studies concerned the open top region due to the
complexity of the analysis in the threshold region.
It is, however, conceivable that the form factor measurements
will be carried out first in the top threshold region.
For this reason it has been demanded for some time that serious
simulation studies on sensitivities to the top form factors
should be performed in the threshold region; it may influence
the energy upgrading program of an $e^+e^-$ linear collider.

\begin{figure}[htpb]
  \begin{center}
\vspace{-1cm}
    \psfrag{(a)}{}
    \includegraphics[height=6cm]{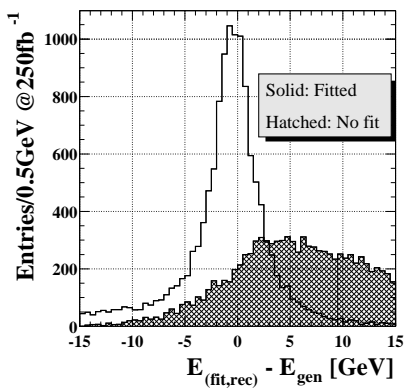}
  \end{center}
  \caption{\small
        Distributions of the difference of the
        reconstructed and generated energies
        of leptonically-decayed $W$ bosons. \cite{ike}}
  \label{EresW}
\end{figure}
To facilitate such a simulation study, a likelihood fitting method
has been devised for kinematical reconstruction of event profiles,
which is particularly tailored to the threshold region \cite{ikematsu}.
The impact of using this method
in the kinematical reconstruction can be seen in the reconstruction
of the energy of leptonically decayed $W$'s.
Fig.~\ref{EresW} shows the difference of the reconstructed $W$ energies
with and without applying the likelihood fitting method.
Without the likelihood fitting, reconstructed $W$ energy
tends to be larger than the generated value.
This is because all the missing momenta, which come from
cascade decays in the
$b$ jets, are mis-assigned to the neutrino in the $W$ decay.
On the other hand, we see a significant recoverly of the generated
energy after the likelihood fitting.
Thus,
we expect that the new kinematical fitting method would be useful
e.g.\ in the measurements of the top decay form factors.

\section{Jet momentum distributions in a flux-tube model}

A flux-tube model has been proposed to predict the 
jet fragmentation properties \cite{lee}.
It is assumed that the number of jets in a process
is determined by the
hard processes in perturbative predictions, whereas all the 
fragmentation inside each jet is predicted by
the flux-tube model in momentum space, without any input from
perturbative QCD.
Based on several assumptions the model is capable of predicting the
momentum distribution of hadrons in each jet.
The distribution for the two-jet case has been reported in this
workshop.
It is important to test the prediction by comparison to 
experimental data.

\section{Future programs}

As seen above, 
interesting and important studies on top and QCD physics
have been performed since the last LC Workshop.
Yet, there remain tasks that have to be done before actual
operation of a next-generation $e^+e^-$ linear collider.
Among various tasks which will hopefully be done 
until the next LC Workshop, I would
like to stress the following ones:
\begin{itemize}
\vspace{-2mm}
\item
Analyses of the top quark form factors both in the open top
and threshold regions, together with
development of new techniques.
\vspace{-2mm}
\item
More accurate theoretical predictions for the relation between
the top quark $1S$ mass and the $\overline{\rm MS}$ mass,
as well as for the normalization of the top production cross section
in the threshold region.
\vspace{-2mm}
\item
Analyses of systematic uncertainties in extracting the top quark mass from
the top--jet invariant mass.
\end{itemize}

\end{document}